\documentstyle[preprint,prl,aps]{revtex}
\begin{document}
\draft

\title{Is a doped 'Kondo' insulator different from doped Silicon? }
\author{J. F. DiTusa$^{1}$, K. Friemelt$^{2}$, E. 
Bucher$^{2}$, G. Aeppli$^{3}$, and A.P. Ramirez$^{4}$}

\address{$^{1}$Department of Physics and Astronomy, Louisiana State 
University, Baton Rouge LA 70803, $^{2}$University of Konstanz, 
Fakultat fur Physik, Postfach 5560, D-78434 Konstanz, 
$^{3}$NEC, 4 Independence Way, Princeton, NJ 08540,
$^4$Bell Laboratories, Lucent Technologies, Murray Hill, NJ 07974,
} 
 
\date{\today}

\maketitle

\begin{abstract}
	We have observed the metal--insulator transition in the strongly
correlated insulator FeSi with the chemical substitution of Al at the Si site. 
The magnetic susceptibility, heat capacity, and field--dependent 
conductivity are measured for Al concentrations ranging from 0 to 0.08. For
concentrations greater than 0.01 we find metallic properties quantitatively
similar to those
measured in Si:P with the exception of a greatly enhanced quasiparticle 
mass. Below 2 K the temperature and field--dependent conductivity 
can be completely described by the theory of disordered Fermi Liquids.

\end{abstract}


Lightly doped semiconductors and insulators lie at the heart of both 
microelectronic technologies and modern condensed matter physics. The most 
prevalent and best characterized of these systems continues to be carrier
doped Si, a simple band insulator. The electronic and magnetic properties 
of carrier doped Si
near the metal--insulator (MI) transition have been the subject of a large 
and growing literature where it is well established that these properties are
determined by the disorder and
electron--electron (e-e) interactions\cite{rosenbaum}. Investigations 
of the MI 
transition where the Coulomb interaction is more significant, such as
the Mott--Hubbard systems V$_{2-x}$O$_3$,
La$_{2-x}$Sr$_x$CuO$_4$, and Ni(S,Se)$_2$\cite{mcwhan}
have revealed interesting magnetic and superconducting ground states. 
A key feature separating Si
from the Mott--Hubbard insulators is that whereas Si is an insulator by
virtue of its band structure, the latter are band metals which are
insulating because of Coulomb interactions. Recently, another class of insulators
has emerged, namely insulators correctly predicted by band theory to be 
insulating\cite{mattheiss}
yet which also display optical and magnetic properties
not understandable using band 
theory\cite{jaccarino}.  
The extent to which such
insulators, frequently labeled as 'strongly correlated' or 'Kondo'
insulators, are fundamentally different from conventional insulators and 
semiconductors is unclear\cite{aepplirev}. In the present paper, 
we present strong evidence that one
popular strongly correlated insulator, FeSi, is actually a renormalized version of
Si in the same sense that the heavy
fermion metals are Fermi liquids with vastly renormalized band masses.
More specifically, we show that chemical doping by Al substitution onto the
Si sites yields a metal which is very similar to Si doped just beyond its
metal-insulator transition. The only obvious distinction is that in
the metal derived from FeSi, the quasiparticle
mass is greatly enhanced relative to that in doped Si.  Thus we have carrier
doped a strongly correlated insulator 
through the MI transition to obtain a disordered Fermi liquid ground state
with a large carrier mass - a heavy fermion metal. Our data represent
convincing support for the proposition 
that while heavy fermion metals and strongly correlated insulators have 
peculiar temperature-dependent properties, their ground states remain 
the well understood Fermi liquids and band insulators.

	The samples were either polycrystalline pellets or small bars cut from 
large Czochralski grown single crystals\cite{kloc}. Polycrystalline samples 
were produced from high purity starting
materials by arc-melting in an argon atmosphere. To improve sample homogeneity
they were  annealed for one
week at 1000$^o$ C in evacuated quartz ampoules.
Powder X-Ray spectra showed the samples to be
single phase with a lattice constant linearly dependent
on Al-concentration as can be seen in Fig. 1a. 
The linearity demonstrates that Al successfully 
replaces Si in the concentration range investigated. Energy dispersive X-ray 
microanalysis (EDX) yielded results consistent with the nominal Al concentration. 
The electrical conductivity was measured on rectangular 
samples with thin Pt wires attached to four
contacts made with silver paste. We collected transverse
magnetoconductance (MC) and Hall effect measurements at 19 Hz 
using lock-in techniques. Finally,
magnetic susceptibility ($\chi$) measurements were made between 1.7 and 400 K
in fields of 0.1 T in a SQUID magnetometer and
the specific heat was established using a standard semi-adiabatic 
heat pulse technique.

The defining feature of an MI transition is the appearance of 
conduction at low temperatures. The carriers manifest themselves in
a variety of standard quantities, including not only the temperature-dependent
conductivity itself, but also the Hall effect, Pauli susceptibility,
specific heat, and MC.  Together, these quantities define the 
most fundamental parameters, namely the carrier density ($n$), sign 
(hole or electron), effective  mass ($m^*$),
elastic scattering length, and e-e interaction strength,  of the
nascent metal. Arguably the most fundamental are the carrier density and sign. 
Our Hall effect data  demonstrate
that Al doping onto the Si sites in FeSi does what is naively expected and
of course also happens when Al is added to pure Si, namely donate one hole per added
Al. Fig. 1(a) shows the corresponding results for $T=1.7 K$. Thermoelectric
effects also depend on the carrier sign, and we have accordingly measured the 
Seebeck coefficient for some of our samples.
The  doped samples display a positive coefficient over the
entire  30 to 400K measurement interval, while in agreement with Wolfe\cite{wolfe}, 
the pure samples yield sign changes at temperatures above and below a maximum at 50K.

After the carrier density and sign, the next basic information required about
systems which undergo MI transitions is
whether other types of order, e.g. superconducting, charge, or magnetic, 
appear nearby. If this were so, the associated fluctuations could contribute  
to scattering of the carriers and so affect all transport 
and thermodynamic properties.
For FeSi$_{1-x}$Al$_x$, we have identified no finite temperature jumps or other 
singularities in any of our transport or thermodynamic data. In addition, we 
used low temperature specific heat and $\chi$ to search for 
unusual fluctuations at low temperatures. Of particular significance 
is that while there is generally a Curie--Weiss--like rise in $\chi'$ at low $T$, 
the amplitude of this rise appears uncorrelated with composition, and 
corresponds to 1\% of $S=1/2$  and $g=2$ impurities per formula unit.

Having established that Al doping introduces carriers into FeSi$_{1-x}$Al$_x$ 
while not leading to any instability such as magnetic order, 
we turn to a direct search for the MI transition in FeSi$_{1-x}$Al$_x$.
Fig. 2b shows $\sigma$ of pure and Al doped samples. From 250 
to 100 K a thermally activated form,
$\sigma = \sigma_a e^{-\Delta_g /2k_BT}$ with $\Delta_g = 680$ K, characterizes the 
pure sample.
For temperatures between 4 and 100 K, a variable range hopping form
describes the 
data well\cite{friemelt,hunt}. Comparison of the data above 200 K for the pure 
and doped materials 
reveals only a slight change with Al doping. Thus the thermally activated carriers from
the FeSi valence band dominate the transport here. Furthermore,
the small change in room temperature $\sigma$ is consistent with only a small 
loss of carrier mobility.
In contrast, the low temperature $\sigma$ (Fig. 1c) shows a 
dramatic 
change with Al substitution, including metallic behavior for $x\geq 0.01$. 
Assuming that every Al dopant creates one carrier, the critical concentration 
($n_c$) is between $2.2\times 10^{20}$ and 
$4.4\times 10^{20}$ cm$^{-3}$ ($n_c = 3.3 (\pm 1.1)\times 10^{20}$).

Fig. 2c shows that the effect of doping on $\chi$
mirrors the effect of doping on $\sigma$. Specifically, $\chi(T)$, like 
$\sigma(T)$
is the superposition of  a contribution
inherited from the insulator, and a doping-induced contribution.
The most prominent feature of $\chi'$ for the insulator is the well-known
hump above room temperature. The hump is due to the  
appearance of magnetic moments on warming and
can be described using the activated form, 
($\chi(T) = (C/T) e^{-\Delta_{\chi} / k_BT}$) with a Curie
constant $C = 1.9$, which is that 
associated with $g=2$ and $s=3/2$, and -most importantly -a gap $\Delta_{\chi} = 680 K$
identical to the transport (conductivity) gap\cite{jaccarino}. Fig. 2c 
also shows 
clearly the Curie--Weiss--like low temperature term which  was discussed above and   
which is common to all of our samples, doped and undoped.  What is not common to
all samples, and indeed grows with doping, is an essentially
T-independent offset ($\delta \chi$) which is most likely the Pauli term derived from
hole doping. Fig. 1b shows how doping causes the amplitude of this term to rise
from a value consistent with zero in the pure compound. 

If our interpretation of the doping-induced T-independent offset
as a Pauli term is correct, 
 there should also be a linear-in-$T$ ($\gamma T$)
contribution to the specific
heat. We have checked this for some of our doped samples. The
inset to Fig. 2c shows that $\gamma$ is much larger in two doped boules than the
$\gamma = 1.8\times10^{-4}J/mole\; Fe\; K^2$ found in a 
nominally pure crystal\cite{hunt}.
The corresponding effective 
mass ($m^*$) of the carriers calculated from free electron theory is 55 $\pm 5$ 
times the 
bare electron mass, which is reduced to 14$\pm 2 m_e$ when the band
degeneracy ($\nu^{\prime}=8$)
is taken into account\cite{mattheiss}. A similar analysis for an $x=0.05$ sample
prepared at a different time yields $m^* = 17 \pm 2$ $m_e$  or 
$4.25\pm 1 m_e$ with $\nu^{\prime} = 8$. 
For comparison $m^*$ from $\chi$ is
$54\pm5 m_e$ which is reduced to $14\pm2 m_e$ by the band degeneracy.
This mass, together with the free electron theory for a parabolic band which
begins to be filled at $n_c$, yields the solid line in Fig.1(b) and
so provides a good description of how $\delta\chi$
varies with $x$. 

 Thus far we have focussed on the gross features
of the MI transition visible at relatively high temperatures.
Because we are dealing with a disordered alloy derived from an insulator
with strong correlation effects, we expect significant 
correlation effects to be visible also in the low-$T$
$\sigma$ and magnetoconductance (MC). Fig. 3 demonstrates that FeSi$_{1-x}$Al$_x$ 
is not
disappointing in this regard. In particular, there is a large contribution
to $\sigma(T,H)$ which is proportional to $\sqrt{T}$ and whose sign can be switched 
by a 16T magnetic field. In addition, there is 
a large MC with a composition-dependent shape. The shape
also varies among samples with the same composition, and seems more
correlated with the amplitude of the Curie--Weiss--like term in $\chi'$ than with 
any other
parameter. Therefore, to understand the behavior of all samples,
one needs to use the theory of Fermi liquids in the presence of ordinary
localizing disorder as well as spin-flip(SF) and spin-orbit(SO) scattering 
terms\cite{lee,rosenbaum}. We have applied this theory to calculate 
the lines through the data
in Fig. 3, and will give further details elsewhere. What is most important, though,
is that for one x=.015 sample, SF and SO processes could be ignored in
a single consistent description of all of our $T$- and $H$-dependent data.
Thus, the moments formed at high temperatures in FeSi$_{1-x}$Al$_x$ appear to be 
completely independent of the Fermi liquid of holes
induced by Al doping.  
 
We have shown that FeSi$_{1-x}$Al$_x$ undergoes a metal-insulator transition as 
a function of $x$. The transition is not accompanied by 
strongly enhanced magnetic fluctuations, as would be expressed 
by a high Wilson ratio ($\chi k_B^2\pi^2 / 3 \gamma \mu_B^2 \simeq 2$ for FeSi), 
and seems unexceptional in all but one
respect, the relatively high effective mass of the carriers. How
unexceptional is clear from Table I , which summarizes our results and compares them 
with those for other materials near their metal-insulator transitions. Even the
magnitudes of the conductivities as a function of both reduced composition $(n/n_c)$
and temperature, as expressed in the coefficient of $\sqrt{T}$ in the low 
temperature data, are similar in FeSi$_{1-x}$Al$_x$ and systems based on the 
classical semiconductors. Parallels exist not only for the low temperature limiting
properties but also for the cross-over to the high temperature regime, which in
the case of FeSi$_{1-x}$Al$_x$ is dominated by the small gap of the insulating parent.
Specifically, Fig. 2a shows the conductivity of Si:P measured by Chapman et 
al.\cite{chapman} 
up to 400 K for samples with $n/n_c$ in the same range as in our measurements for
FeSi$_{1-x}$Al$_x$. It is striking that for both Si:P and FeSi$_{1-x}$Al$_x$, 
the conductivity rises linearly with decreasing $T$ over a large range of $T$, as
of course it does for many high-T$_c$ superconductors and certain rare earth
intermetallics\cite{aronson}. The associated slopes 
$d\sigma / dT$ appear to scale with distance
from the MI transition, i.e. $\sigma(T,n) = \sigma_o - bT(n/n_c)$ with a 
coefficient $b$ which seems to be a property of the alloy series in question. 
From the last column in the table, it is clear that $b$ for FeSi$_{1-x}$Al$_x$
is of the same order as for the doped (band) semiconductors Si:P and Si:As.
At the same time, it is substantially larger than for Si:B and Ge:Sb. What is 
perhaps most striking is that $b$ for FeSi$_{1-x}$Al$_x$, Si:P, and Si:As is of 
the same order as $b$ for the doped Mott--Hubbard insulator and high-T$_c$ 
superconductor La$_{2-x}$Sr$_x$CuO$_4$, which is considerably more famous for its
linearly $T$-dependent resistivity than the classic semiconductors.

	In summary, we have shown how a heavy fermion metal emerges upon doping
the strongly correlated insulator, FeSi. The resulting metals and associated
insulator--metal transition bear an extraordinary and even quantitative
resemblance to those near the 
classic metal--insulator transition in the more conventional insulator Si. Thus,
the strong Coulomb effects present in insulating FeSi serve only to renormalize
the critical concentration $n_c$ and effective carrier masses in the 
metallic phase.

	We thank P.W. Adams, D.A. Browne, Z. Fisk, and D.R. Hamann for 
discussions. JFD acknowledges the support of the Louisiana Board of Regents through 
contract number LEQSF(RF/1995-96)-RD-A-38.

\vfill\eject
\noindent\centerline{
\footnotesize{
\tabcolsep=0.07in
\doublerulesep=0.15in
\begin{tabular}{|llllllccl|} \hline
System  & $E_g$ (eV) & $n_c (10^{18}cm^{-3})$ & $m^* / m_e$ & $k_F\ell / 3$ & 
$\sigma_o (1 /\Omega cm)$ & $\nu$ & $a$ $(\Omega cm K^{1/2})^{-1}$ & 
$b$ ($\Omega cm$ $K$)\\ \hline
FeSi$_{1-x}$Al$_x$ & 0.06 & $330\pm110$ & $15\pm 2$ & $0.23$ & 
$600\pm150$ & $0.85\pm.1$ & $-10.0$ & $-1.10\pm0.2$ \\
Si:P & 1.17 & $3.74$ & $0.26$ & 0.46 & $260\pm30$ & $0.55\pm.1$ &
$-3.0$ & $ -0.41\pm0.07$  \\
Si:B & 1.17 & $4.06$ & $0.38$ & $0.20$ & $152\pm18$ & $0.65\pm.14$
& $-7.0$ & $-0.062\pm0.005$ \\
Si:As & 1.17 & $8.2$ & $0.31$ & $0.34$ & $381$ & $0.64\pm .2$ &
$-11.0$ & $-8\pm2$ \\
Ge:Sb & 0.75 & $ 0.15$ & $0.22$ & $0.51$ & $63\pm14$ & $0.7\pm.2$
& $-12.0$ & $-0.073\pm0.008$ \\
La$_{2-x}$Sr$_x$CuO$_4$ & 1.8 & $260$ & $2$ & $\sim 3$ & { } & { }&
{ } & $-1.5\pm0.3$ \\ \hline
\end{tabular}
}}
\begin{table}
\caption{
Values of the energy gap in the undoped systems ($E_g$), 
critical concentration $n_c$, the effective carrier mass $m^*/m_e$, the 
dimensionless diffusion constant ($k_F\ell / 3$) for $n=2n_c$ calculated 
using the dopant density as the nominal carrier concentration, a valley 
degeneracy of 8 for FeSi, 6 for Si, and 4 for Ge, and the free electron formula
for $\sigma$ {\protect\cite{rosenbaum}} (the value for 
La$_{2-x}$Sr$_x$CuO$_4$ was scaled from a high temperature estimate given in
Ref. {\protect\cite{takagi}}), 
$\sigma_o$, and $\nu$ obtained from fits of the data to the form $\sigma = 
\sigma_o
(n/n_c - 1)^{\nu}$, $a$ from fits to the form $\sigma = \sigma_o + aT^{1/2}$ to the 
low temperature conductivity at zero field, and the parameter b from fits to the 
linear temperature dependent conductivity to the form 
$\sigma(T,n) = \sigma_o(n) + b(n/n_c)T$. Data for Si:P were taken from ref.
{\protect\cite{rosenbaum,chapman,putley}}, for Si:B from ref. 
{\protect\cite{rosenbaum,chapman,putley}}, for Si:As from  ref. 
{\protect\cite{putley,newman}}, for Ge:Sb from ref. 
{\protect\cite{putley,thomas}}
and for La$_{2-x}$Sr$_x$CuO$_4$ from ref. {\protect\cite{takagi,cooper}}.
	}
\label{table1}
\end{table}
\begin{figure}
\caption{
a) Lattice constant of FeSi$_{1-x}$Al$_x$ vs. nominal Al concentration
($\bullet$ left hand axis) and nominal carrier concentration from measurements
of Hall voltage ($\Box$ right hand axis).
b) Change in $\chi$ with $x$, taken as the average $\chi$ between 80 and 120 K. 
Solid line is best fit to the form $\delta\chi = c(n - n_c)^{1/3}$ with 
$c = 1.44\mu_B^2\nu^{\prime\; 2/3} m^*/\hbar^2\pi^{4/3}$ where $\nu^{\prime}$ is 
the valley degeneracy
taken as 8{\protect\cite{mattheiss}}. The best fit corresponds to a carrier 
mass of $14m_e$.
(c) The low temperature conductivity vs.\ nominal Al concentration ($\bullet$).
Si:P data ($\Box$ left hand axis only){\protect\cite{thomas,rosenbaum3}} is 
plotted for comparison with $n_c = 3.74\times10^{18}$. $\sigma_{min}$ (left hand 
axis) is the Mott minimum conductivity defined as $\sigma_{min}= 0.05 e^2 / 
\hbar d_c$ with $d_c = n_c^{-1/3}$.
Solid line represents a fit of FeSi$_{1-x}$Al$_x$ data up to the 
concentration where the Ioffe--Regel condition is violated ($k_F\ell\sim 2$, 
$x\le 0.045$) to $\sigma_{LT} =
\sigma_o * (n/n_c -1)^{\nu}$ with best fit corresponding to $\nu = 0.9\pm 0.1$
and $\sigma_o = 190 \pm 40$. Dashed dotted 
line represents best fit of the Si:P data to the same form ($\nu=0.55$).
	}
\label{fig2}
\end{figure}

\begin{figure}
\caption{
a) $\sigma(T)$ for Si:P with P concentrations of $2.7\times10^{19}$ 
($\bullet$), $1.6\times10^{19}$ ($\ast$), $1.1\times10^{19}$ ($\diamond$), 
$7.8\times10^{18}$ ($\triangle$), $4.9\times10^{19}$ ($\Box$), $2.8\times10^{18}$
($\times$) data of P.W. Chapman et al.{\protect\cite{chapman}}.
b) $\sigma(T)$  for FeSi$_{1-x}$Al$_x$ with $x$ of
0.0 single crystal ($\bullet$), 0.0 (solid line), 0.005 (dashed line), 
0.01 ($\triangleleft$), 0.015 ($\circ$), 
0.015 ($\ast$), 0.025 single crystal ($\bigtriangledown$), 0.025 ($\bigodot$),
0.035 (dashed dotted line), 0.045 ($\times$), 0.05 ($\diamond$), 0.055 
($\triangleright$), 0.06 ($\triangle$), 0.07 ($+$), 0.08 ($\Box$). 
c) $\chi(T)$ for FeSi$_{1-x}$Al$_x$ with symbols same as in a). 
Inset: $C(T) / T$ plotted as a function of $T^2$ for $x=0.015$ ($\ast$) and 
$x=0.025$ ($\bigodot$).	}
\label{fig1}
\end{figure}

\begin{figure}
\caption{
a) $\Delta\sigma$ vs. $T^{1/2}$ for FeSi$_{0.985}$Al$_{0.015}$ in 0 and 
16 T. 
Lines represent best fits to the low-$T$ data to
$\sigma(T)\propto T^{1/2}$.
b) Magnetoconductivity at 0.290 K for FeSi$_{1-x}$Al$_x$ with $x=0.015$ 
($\circ$), $x=0.015$ ($\ast$), $x=0.025$ ($\bigodot$), and $x = 0.05$ ($\diamond$).
Solid line through data in both (a) and (b) for $x=0.015$ ($\circ$) 
is calculated from
theory {\protect\cite{lee}} with the gyromagnetic ratio for the carriers of 
2.75. Other lines represent fits to theory which includes effects
of spin orbit or spin flip scattering. }
\label{fig3}
\end{figure}

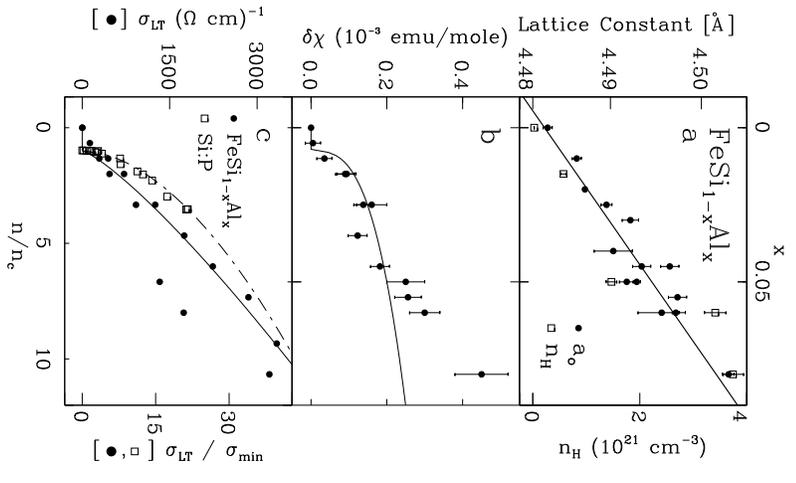
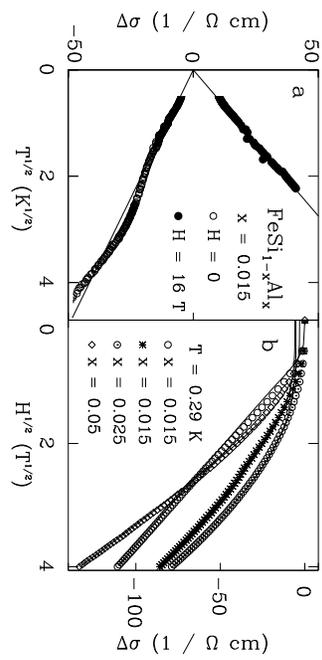
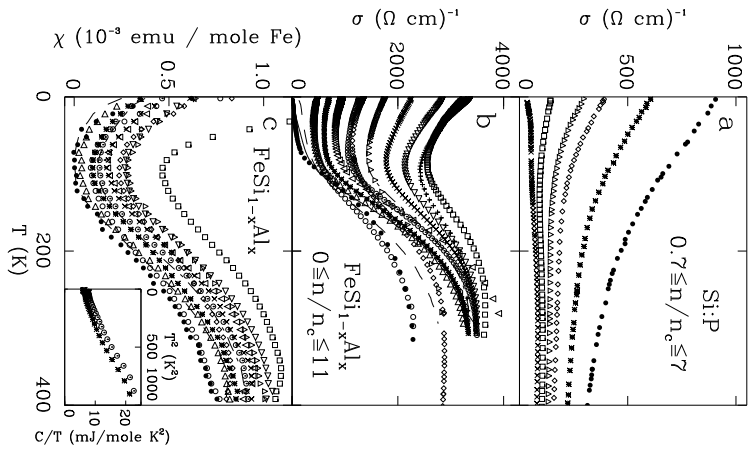
\end{document}